\documentclass{article}
\usepackage[utf8]{inputenc}
\usepackage{amsmath,amssymb,amsfonts}
\usepackage{graphicx}
\usepackage{hyperref}
\usepackage{xcolor}
\usepackage{listings}
\usepackage{indentfirst}
\usepackage{float}
\usepackage{booktabs}
\usepackage{caption}
\usepackage{array}
\usepackage{subcaption}
\usepackage{geometry}
\usepackage{rotating}    

\lstdefinestyle{mystyle}{
    backgroundcolor=\color{codegray},   
    commentstyle=\color{codegreen},
    keywordstyle=\color{codeblue},
    numberstyle=\tiny\color{codegray},
    stringstyle=\color{codered},
    basicstyle=\ttfamily\footnotesize,
    breakatwhitespace=false,         
    breaklines=true,                 
    captionpos=b,                    
    keepspaces=true,                 
    numbers=left,                    
    numbersep=5pt,                  
    showspaces=false,                
    showstringspaces=false,
    showtabs=false,                  
    tabsize=4
}

\title{Heterogeneous Trader Responses to Macroeconomic Surprises: A Simulation of Order‑Flow Dynamics}

\author{Haochuan(Kevin) Wang\\
\texttt{haochuanwang@uchicago.edu}\\
University of Chicago}

\date{March 2025}

\begin{document}

\maketitle

\begin{abstract}
\noindent
Understanding how market participants react to shocks like scheduled macroeconomic news is crucial for both traders and policymakers.  We develop a calibrated data‐generation process (DGP) that embeds four stylized trader archetypes—retail, pension, institutional, and hedge funds—into an extended CAPM augmented by CPI surprises.  Each agent’s order‐size choice is driven by a softmax discrete‐choice rule over small, medium, and large trades, where utility depends on risk aversion, surprise magnitude, and liquidity. We aim to analyze each agent's reaction to shocks and Monte Carlo experiments show that (i) higher‐information, lower‐aversion agents take systematically larger positions and achieve higher average wealth; (ii) retail investors under‐react on average, exhibiting smaller allocations and more dispersed outcomes; and (iii) ambient liquidity amplifies the sensitivity of order flow to surprise shocks.  Our framework offers a transparent benchmark for analyzing order‐flow dynamics around macro releases and suggests how real‐time flow data could inform news‐impact inference.
\end{abstract}

\section{Introduction}
Scheduled macroeconomic announcements—such as the Consumer Price Index (CPI) and FOMC releases—are among the most important days accompanied by high trading flows and volatility spikes. Previous studies document large spikes in trade volume and transient price dislocations following surprises in macro data. At the same time, market participants differ widely in risk tolerance, information access, and execution speed. Yet, despite a growing literature on macro surprises and asset returns, relatively little is known about how heterogeneous traders’ behaviors interact to produce the observed cross‐sectional patterns in order flow and reality.

Due to the constraint and limited access, this paper bridges that gap by introducing a simulation‐based framework that tries to explore the behavioral differences in which four trader types—retail investors, pension funds, institutional investors, and hedge funds—choose discrete order sizes via a softmax‐based utility maximization.  We extend the standard CAPM to include both temporary and permanent components of CPI surprises, and we calibrate each agent’s risk aversion and information quality to stylized empirical values.  A Monte Carlo grid search over portfolio weights then yields each trader’s optimal allocation under varying surprise magnitudes and liquidity conditions.  Our contributions are threefold:

\begin{enumerate}
  \item We demonstrate how differences in risk aversion and information translate into systematic differences in trade size and wealth accumulation across agent types.
  \item We quantify the amplifying role of market liquidity in shaping the sensitivity of order flow to macro surprises.
  \item We provide a transparent benchmark model that can be extended and get closer to reality: given observed order‐flow patterns around macro releases, one could infer plausible surprise magnitudes or heterogeneity parameters.
\end{enumerate}

\section{Literature Review}

\subsection{Impact of Macroeconomic Surprises on Trading}
A consistent finding in recent empirical work is that major macroeconomic announcements provoke significant trading activity in financial markets. For example, an extensive study of UK markets by Heinlein and Lepori finds that a “large number of macroeconomic announcements increase trading activity in the stock market,” indicating that news releases lead market participants to revise expectations and trade accordingly. And the same research mentions that while volume surges, price swings are caused by significant amount of trading activities from various agents. 

Beyond volume, researchers have probed how price formation and return patterns reflect the heterogeneous processing of macro news. Han (2025) develops a dynamic noisy rational-expectations model to explain empirical anomalies around announcement days. In the paper, it mentioned that when some investors are better informed than others, informed traders’ forecast revisions only imperfectly trickle into average expectations, allowing “noise” to build up in prices prior to the news. Once the announcement occurs and public information resets beliefs, that accumulated noise gets corrected – resulting in price reversals or unusual return patterns on announcement day. In the later DGP, we will find ways to capture the differences with simulated noises and adjust the scaling parameter on the variance according to various agents.

\subsection{Heterogeneous Portfolio Rebalancing}
When a major macroeconomic surprise happens in the market, investors do not all adjust their portfolios in the same way. A growing number of recent research show how heterogeneity in beliefs and constraints leads to different portfolio rebalancing in response to macro news. One important insight is that announcements often synchronize certain trading actions by resolving uncertainty and reducing disagreement. Ying (2021) documents this phenomenon around U.S. monetary policy announcements: using high-frequency equity and options data, he finds that the open interest in index options drops by over 50 \(\%\) on FOMC announcement days, implying that a wave of investors are unwinding their pre-positioned trades once the news is publicly available.

Another group of researchers examines how investors move money between asset classes (stocks, bonds, etc.) after macro shocks – and finds that heterogeneity in investor type is crucial. Lu and Wu (2023) focus on institutional investors who allocate across multiple asset classes and propose that their rebalancing activity is a key driver of stock market reactions to interest-rate surprises. The intuition is that certain institutions will sell equities and buy bonds when rates rise unexpectedly, to maintain portfolio targets or exploit higher yields, whereas other investors like equity-focused retail traders may not respond to substituting between asset classes. This heterogeneous behavior will further create an uneven impact: stocks heavily held by multi-asset allocators are more sensitive to the rebalancing sell-off than stocks. However, this is a more complicated set of questions that need to be investigated further and can't be captured with a simpler DGP setting.

\subsection{Differences Between Trader Types}
Perhaps the most practical manifestation of heterogeneous responses to macro news is the contrast between retail investors and institutional investors. A number of post-2020 studies have zeroed in on how these trader groups differ in anticipation of and reaction to economic announcements. Evidence from stock markets suggests a clear divergence in behavior before the news breaks. Jia et al. (2023) analyze proprietary Chinese stock exchange data around a high-profile monthly macro announcement. They find that institutional investors actively de-risk ahead of the announcement: on average, institutions cut their overall stock exposure in the days leading up to the news, while selectively tilting into smaller-cap, riskier stocks to seek higher potential upside. 

And interestingly, many individuals see an asset’s rally on good news and decide to bet on a reversal, rather than update their fundamental outlook. The authors demonstrate that this behavior is predictable and uninformed: simple trading strategies that take the opposite side of aggregate retail order flow around news events are profitable, indicating that retail trades systematically mis-react to macro news. This also reinforces the hypothesis that different market participants have differential abilities to interpret public information papers and institutional investors typically have better models and expertise to quickly digest an employment report or GDP print, whereas retail traders may be swayed by heuristics, recent price movements, or media sentiment. As historical trades and order data are highly restricted by market makers and exchanges, there are limited amounts of publicly available data so we decided to analyze these hypotheses and claims regarding different investors' reactions to shocks with simulated data. All the literature provides us a theoretical background for the parameter tuning for the DGP.

\section{Data Generation Process}

\subsection{Market Participants Heterogeneity}
We classify traders into four types with fixed risk aversion levels: Retail, Pension Funds, Institutional Investors, and Hedge Funds. Risk appetite and market information levels increase progressively across these types. Hedge funds, with the highest risk tolerance and better information access, trade more aggressively on news, placing larger orders. In contrast, retail trade less frequently and in smaller quantities.

We classify traders into four archetypes—Retail, Pension Funds, Institutional Investors, and Hedge Funds—each with fixed risk aversion levels and distinct information sets. This stratification is consistent with the empirical finance literature, where market participants are observed to differ systematically in both preferences and sophistication:

Retail investors: Typically more risk-averse, with limited access to information and execution technologies. Empirical evidence (e.g., Zhu 2023, JPMorgan 2023) shows retail flows are more dispersed, less sensitive to fundamentals, and often contrarian around announcements.



\subsection{Events Expectations and Surprises}

Scheduled macroeconomic announcements, such as CPI releases, are preceded by consensus forecasts published by data providers like Bloomberg and Refinitiv. Markets react not to the level of the data but to the \emph{surprise} component---defined as the difference between realized and expected values:

\[
\text{Surprise} = \text{Actual} - \text{Expected}.
\]

For example, if CPI is expected at 3.3\% and arrives at 3.5\%, the surprise is +0.2 percentage points. Importantly, the sign of a surprise is state-dependent. For instance positive CPI surprise in this case signals stronger inflationary pressure, raising the probability of tighter monetary policy, higher yields, and weaker equities.Conversely, a negative CPI surprise suggests easing inflationary pressure, supporting risk assets but hurting inflation hedges.

\subsection{Asset Classification}

For tractability, the DGP begins with two representative assets one is Risk-free asset where offering stable returns and in this case insensitive to interest rate surprises like a short term treasury. And another is Inflation-sensitive risky asset: A stylized equity or bond index that reacts directly to inflation news.This binary setup balances simplicity with interpretability: traders choose between a safe allocation and an asset whose payoff is linked to macroeconomic conditions. While real markets feature a richer asset space (equities, bonds, commodities, derivatives), a two-asset system suffices to highlight cross-sectional differences in trader behavior and utility-driven rebalancing.

\subsection{Extended CAPM Framework}
Building on the Capital Asset Pricing Model (CAPM), we aim to measure how asset returns are determined not only by traditional market risk but also by unexpected macroeconomic shocks. In the standard CAPM, the excess return on an asset is explained solely by its sensitivity to the market premium. While, in this project, asset returns also depend on macroeconomic shocks, which decomposed into temporary and permanent components. 
\begin{equation}
R_i - R_f = \alpha_i + \beta_i (R_m - R_f) + \gamma_i S + \delta_i S^P + \epsilon_i,
\end{equation}

Here, \(\beta_i (R_m - R_f)\) is the traditional market premium while \(\beta_i\) measures the asset's sensitivity to overall market movements and often note as market premium.
\( \gamma_i S\) captures the asset's response to the overall CPI surprise. And \(\delta_i S^P \) measures the permanent component of the CPI surprise. In this case it will be the shock premium. And notice there might be structural changes in the market condition after some huge shocks, we then decide to use\(\delta_i\) to measure how much of the asset’s return is affected by these long lasting changes, ie. huge inflation rate that could alter the entire market. And the \(\alpha_i\)
here accounts for asset-specific characteristics not captured by other factors.

\subsection{Utility Function}
In our simulation, each trader selects an order size from a discrete set (e.g., \textit{small}, \textit{medium}, \textit{large}). The instantaneous utility for trader \( k \) choosing order size \(\ell \in \{\text{small},\,\text{medium},\,\text{large}\}\) is modeled as:

\begin{equation}
U_{k}(\ell) \;=\; c_{0,\ell} \;+\; c_{\text{RA},\ell}\,\mathrm{RA}_{k} \;+\; c_{\text{sur},\ell}\,\bigl|\mathrm{Surprise}\bigr| \;+\; c_{\text{liq},\ell}\,\mathrm{Liquidity} \; 
\end{equation}

Here we have \(c_{0,\ell}\) is a baseline utility term for order size \(\ell\); 
\(\mathrm{RA}_{k}\) is trader \(k\)'s risk aversion level; \(\bigl|\mathrm{Surprise}\bigr|\) is the magnitude of the CPI surprise; \(\mathrm{Liquidity}\) is a level of market liquidity.

\subsection{Discrete Choice via Softmax}
Given we can measure the utility of each trader's transaction, we further construct the choice probabilities as a softmax function for each trader. 
\begin{equation}
P(\ell \mid k) \;=\; \frac{\exp\bigl(U_{k}(\ell)\bigr)}{\displaystyle \sum_{\ell' \in \{\text{small},\,\text{medium},\,\text{large}\}} \exp\bigl(U_{k}(\ell')\bigr)}.
\end{equation}

Here the denominator is the sum of all probabilities, which sum up to 1. And exponentiating the utility magnifies differences, meaning a higher utility leads to a substantially higher probability of the trader \(k\) choose to trade size \(l\). In parallel to the discrete choice of the order size, we capture how traders allocate their wealth between assets and in which order quantity

\subsection{Constant Relative Risk Aversion Utility Maximization}

\begin{figure}[H]
    \centering
    \includegraphics[width=0.5\linewidth]{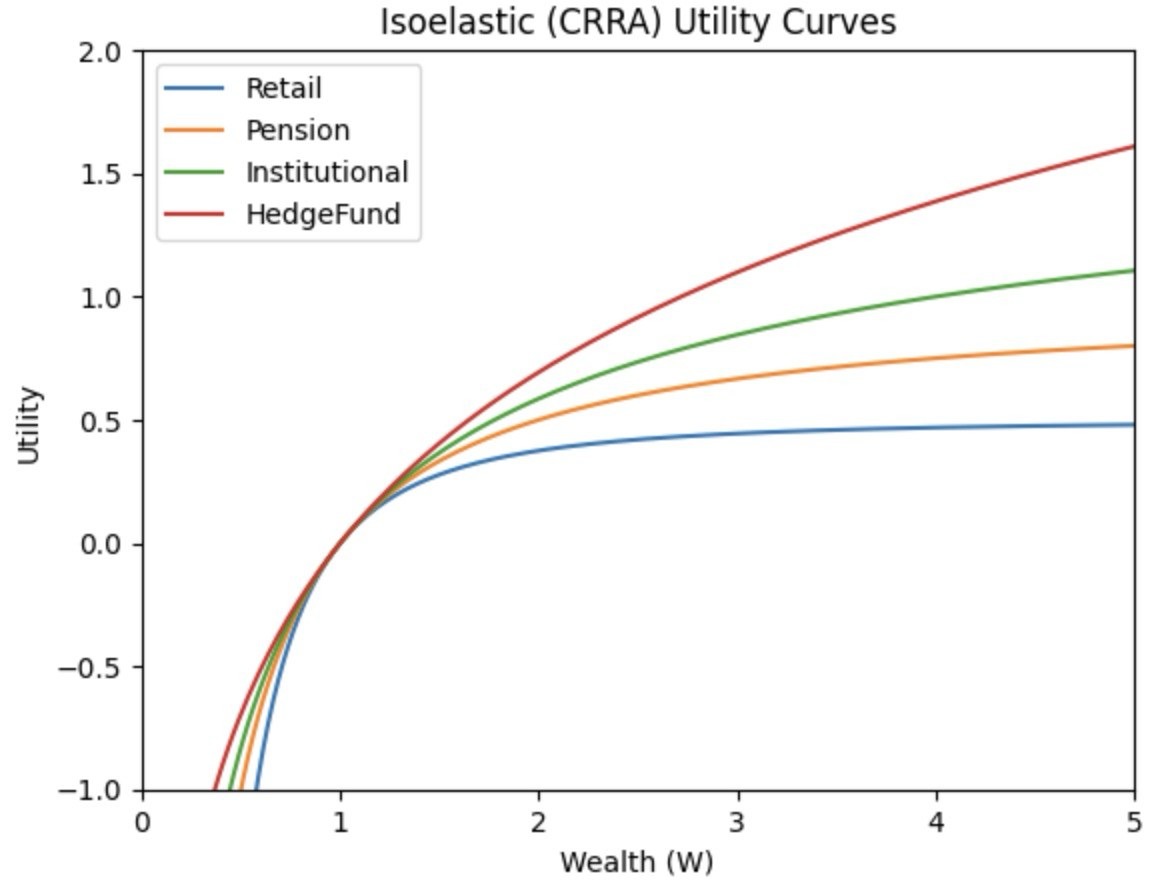}
    \caption{Isoelastic Utility}
    \label{fig:enter-label}
\end{figure}

In the DGP, we assume that each trader’s risk aversion parameter, \(\gamma_k\), is constant over multiple time periods. In the graph we assume Retail Trader has the highest risk aversion \(\gamma_k = 3\), Pension has \(\gamma_k = 2\), Institution has \(\gamma_k = 1.5\), and Hedge Funds has \(\gamma_k = 1\). Traders maximize the expected utility of their final wealth, where utility is given by the isoelastic (CRRA) function:

\begin{equation}
\max_{0 \le x \le 1} \quad \mathbb{E}\bigl[\,U\bigl(W_{\mathrm{final}}\bigr)\bigr]
\quad \text{where} \quad
U\bigl(W_{\mathrm{final}}\bigr) \;=\; \frac{W_{\mathrm{final}}^{\,1-\gamma_{k}}-1}{\,1 - \gamma_{k}\,}.
\end{equation}

with final wealth defined as:
\[
W_{\mathrm{final}} \;=\; W_{0}\,\Bigl[(1 - x)(1+R_f) + x(1+R_r)\Bigr].
\]

Since the \(W_{0}\) is the initial wealth, we can factor out \(W_{0}\) from the utility expression:

\[
\Pi(x,R_r) 
\;=\; (1 - x)\,(1 + R_f) \;+\; x\,(1 + R_r),
\]
Then we substitute into the utility function
\[
U\bigl(W_{\mathrm{final}}\bigr)
\;=\;
\frac{\bigl(W_0\,\Pi(x,R_r)\bigr)^{1-\gamma_k} - 1}{\,1 - \gamma_k\,}
\;=\;
\frac{\,W_0^{\,1-\gamma_k}\,\Pi(x,R_r)^{\,1-\gamma_k} - 1\,}{\,1 - \gamma_k\,}.
\]

Because \(W_0^{1-\gamma_k}/(1 - \gamma_k)\) is a constant factor, solving for the optimal \(x\) is equivalent to
\[
\max_{0 \,\le\, x \,\le\, 1}
\;\; \mathbb{E}\bigl[\Pi(x,R_r)^{\,1-\gamma_k}\bigr]
\;=\;
\max_{0 \,\le\, x \,\le\, 1}
\;\; \mathbb{E}\!\Bigl[\bigl((1 - x)(1 + R_f) + x(1 + R_r)\bigr)^{1-\gamma_k}\Bigr].
\]

In our model, we define \(R_r\) follows a normal distribution with mean 0.1 and standard deviation 0.2.  Because a closed‐form solution for the CRRA utility optimization generally does not exist under this distributional assumption, we thus want to solve this with a numerical (Monte Carlo) approach to approximate the trader’s expected utility and select an optimal allocation.

\textbf{Monte Carlo Approximation}

We implement a discrete grid search over \( x \) to approximate the optimal solution. The interval \([0,1]\) is divided into increments of 0.05, generating candidate allocations. For each \( x \), we draw multiple i.i.d. samples of the risky return \( R_r \) from a normal distribution with mean 0.1 and standard deviation 0.2, adjusting the standard deviation based on each trader’s information level (more informed traders have lower noise).  

For each draw, we compute the final wealth \( W_{\mathrm{final}} \) and evaluate the trader’s CRRA utility \( U(W_{\mathrm{final}}) \). We then estimate \( \max \mathbb{E}[U(W_{\mathrm{final}})] \) by averaging over all samples.  

An alternative approach is to optimize the expectation under exponentiation, leveraging the analytical result for normally distributed variables:  
\[
E[e^X] = e^{\mu + \sigma^2/2}.
\]
However, directly substituting \( E[e^X] \) into the utility function may misrepresent risk preferences, as expected utility depends on the full return distribution, and after some testing and exploration, we decided to settle with Monte Carlo for the DGP.

\subsection{Expectations before the DGP}
One of the primary objectives of this project is to explore how differences in trader behavior and information affect portfolio returns. In our framework, we hypothesize that:
\begin{itemize}
    \item Institutional traders and hedge funds, with professional quantitative agents have more accurate information and lower risk aversion, are likely to achieve higher returns than other traders.
    \item Retail traders, who are less informed, tend to earn the lowest among the four types of traders.
    \item Pension funds are more conservative than institutional traders and hedge funds but are more informed than retail traders, they may choose safer assets and has better portfolio performance than retail.
\end{itemize}

\section{Discussion on DGP Result}

\textbf{Risky Asset Allocation by Periods}

\begin{figure}[H]
    \centering
    \includegraphics[width=0.49\linewidth]{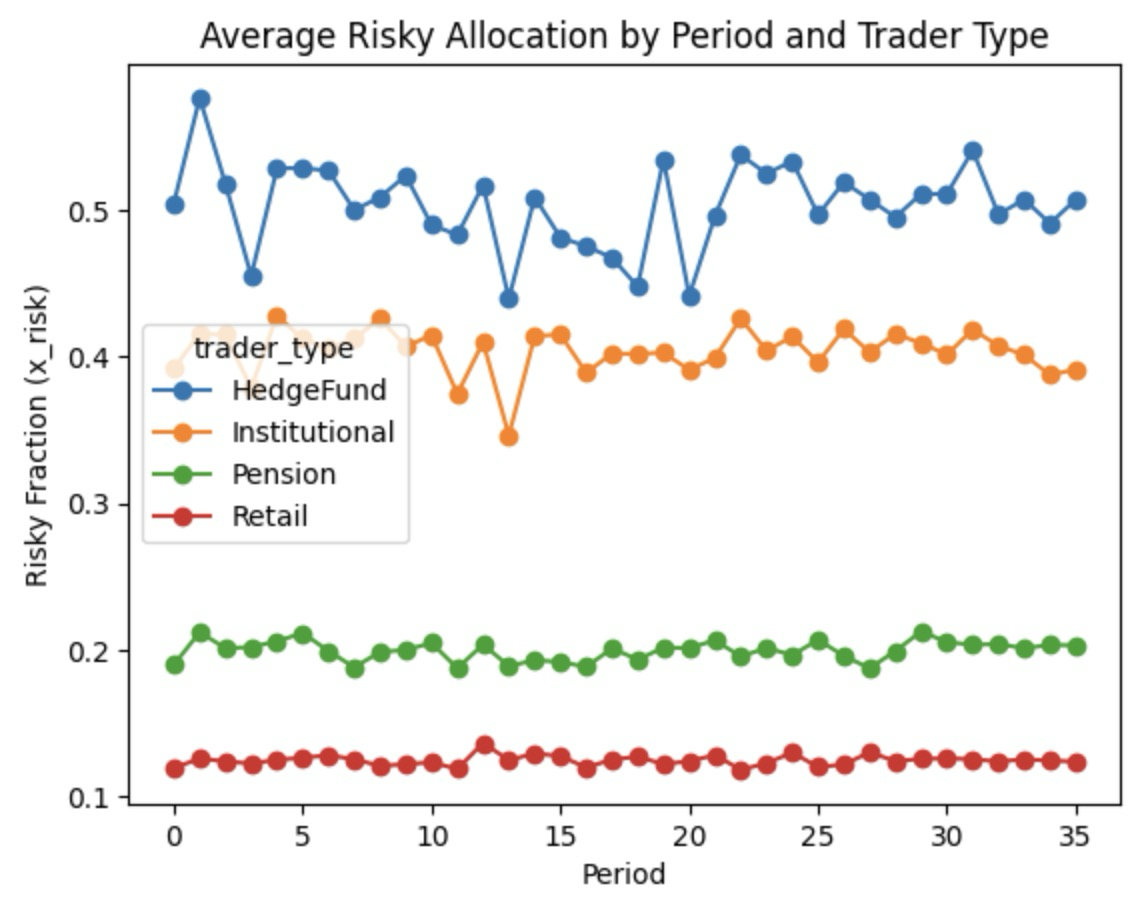}
    \caption{Risky Asset allocation}
\end{figure}

Here, we simulate 500 traders per type across 36 CPI events and analyze their risky asset allocation. The line plot depicts the average fraction of wealth allocated to the risky asset for each trader type. As expected, Hedge Funds allocate the largest fraction, followed by Institutional, Pension, and Retail investors.

In our DGP, Retail traders, having the highest risk aversion and least amount of quality information, perceive greater noise in assessing events, leading to lower and less frequent adjustments in risky asset holdings. This aligns with our assumptions: Hedge Funds exhibit the highest risk appetite, Institutional traders follow, Pension funds are more conservative, and Retail investors are the most risk-averse.

\textbf{Final Wealth Distribution}

\begin{figure}[!h]
    \centering
    \includegraphics[width=0.5\linewidth]{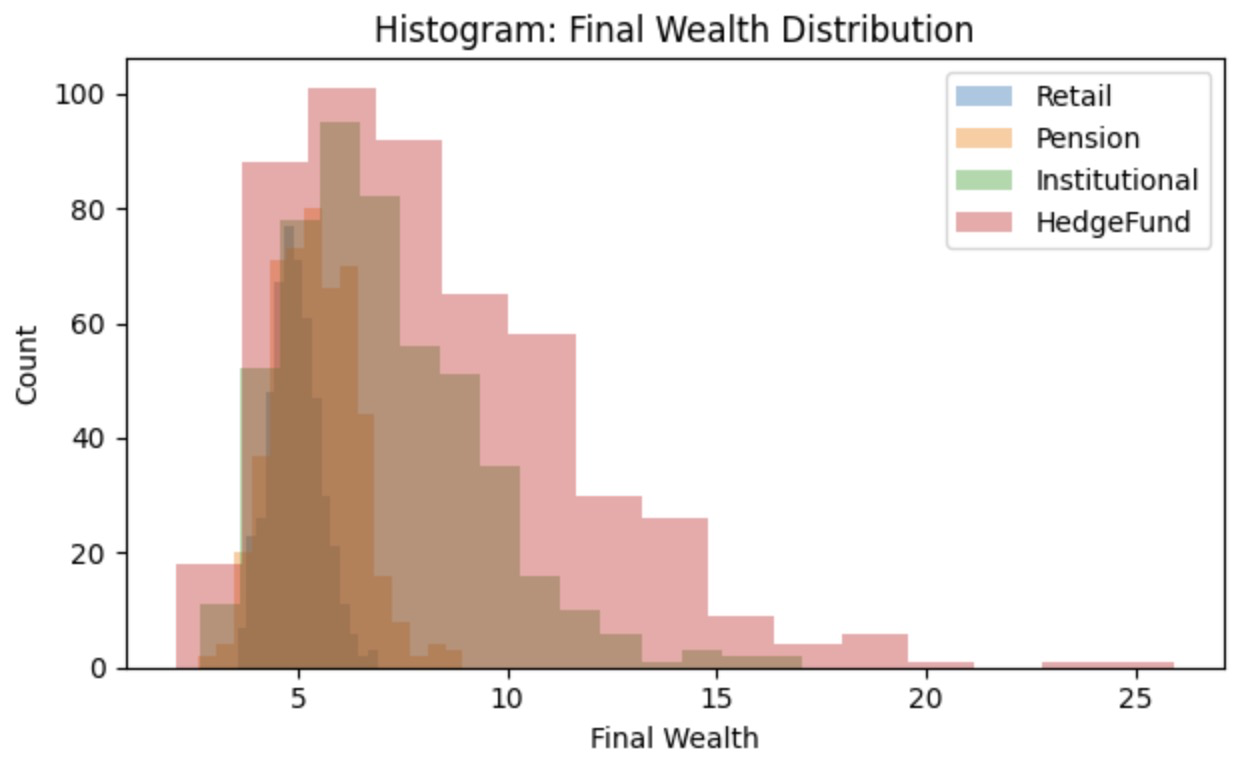}
    \caption{Wealth Distribution}
    \label{fig:enter-label}
\end{figure}

Here, Hedge Funds exhibit the highest exposure to risky assets, resulting in the widest distribution of final wealth. And notice Hedge funds generally are more informed, in the DGP, we normally see them has a better risk appetite and understanding of the risk trade off and making better decisions and the wealth distribution histogram shows evidence to support this point(Figure 3).

Across 500 individuals per trader type, Hedge Funds achieve the highest average wealth over time, reflecting their willingness to take on greater volatility (Figure 4). And with increasing liquidity in the market, there is a higher tendency for all traders to increase the trade, to maximize their overall utility.(Figure 5)\\

\begin{figure}[H]
    \centering
    \includegraphics[width=0.5\linewidth]{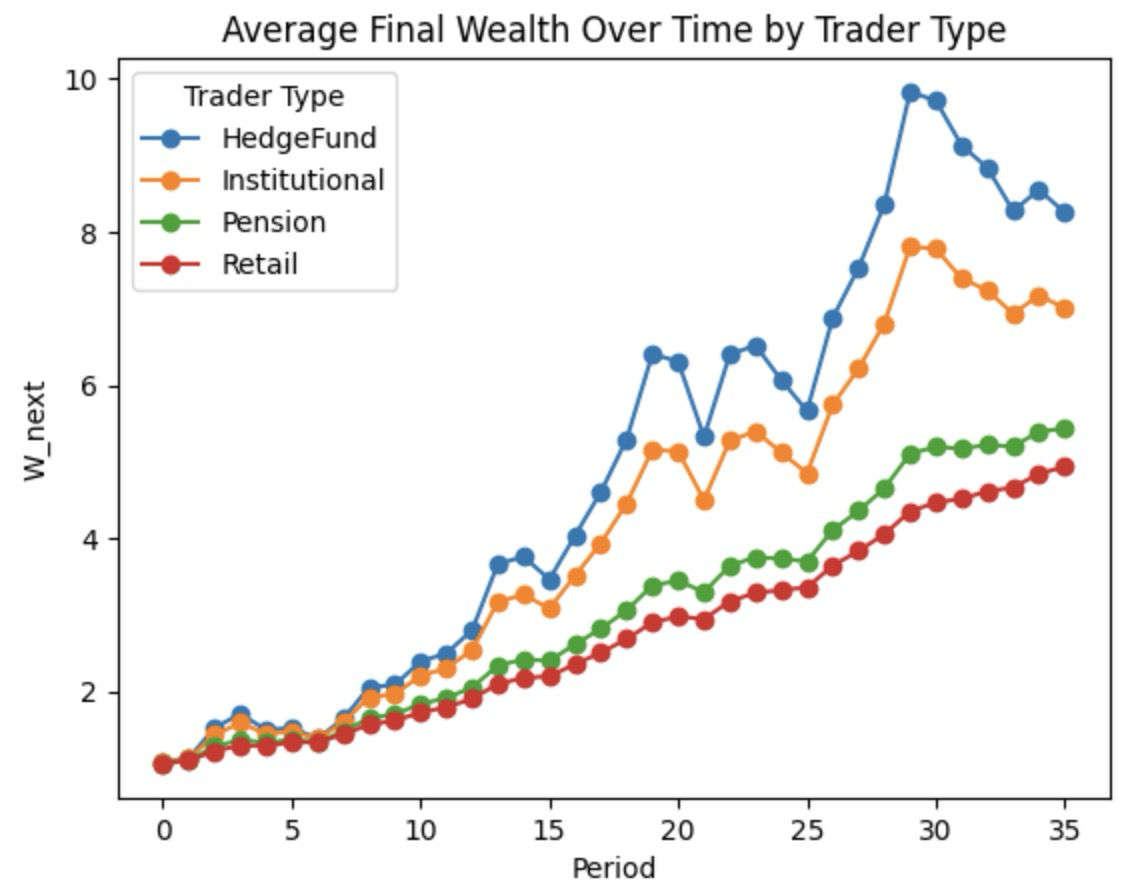}
    \caption{Mean Wealth Accumulation for 500 traders each type}
    \label{fig:enter-label}
\end{figure}




\begin{figure}[H]
    \centering
    \includegraphics[width=0.5\linewidth]{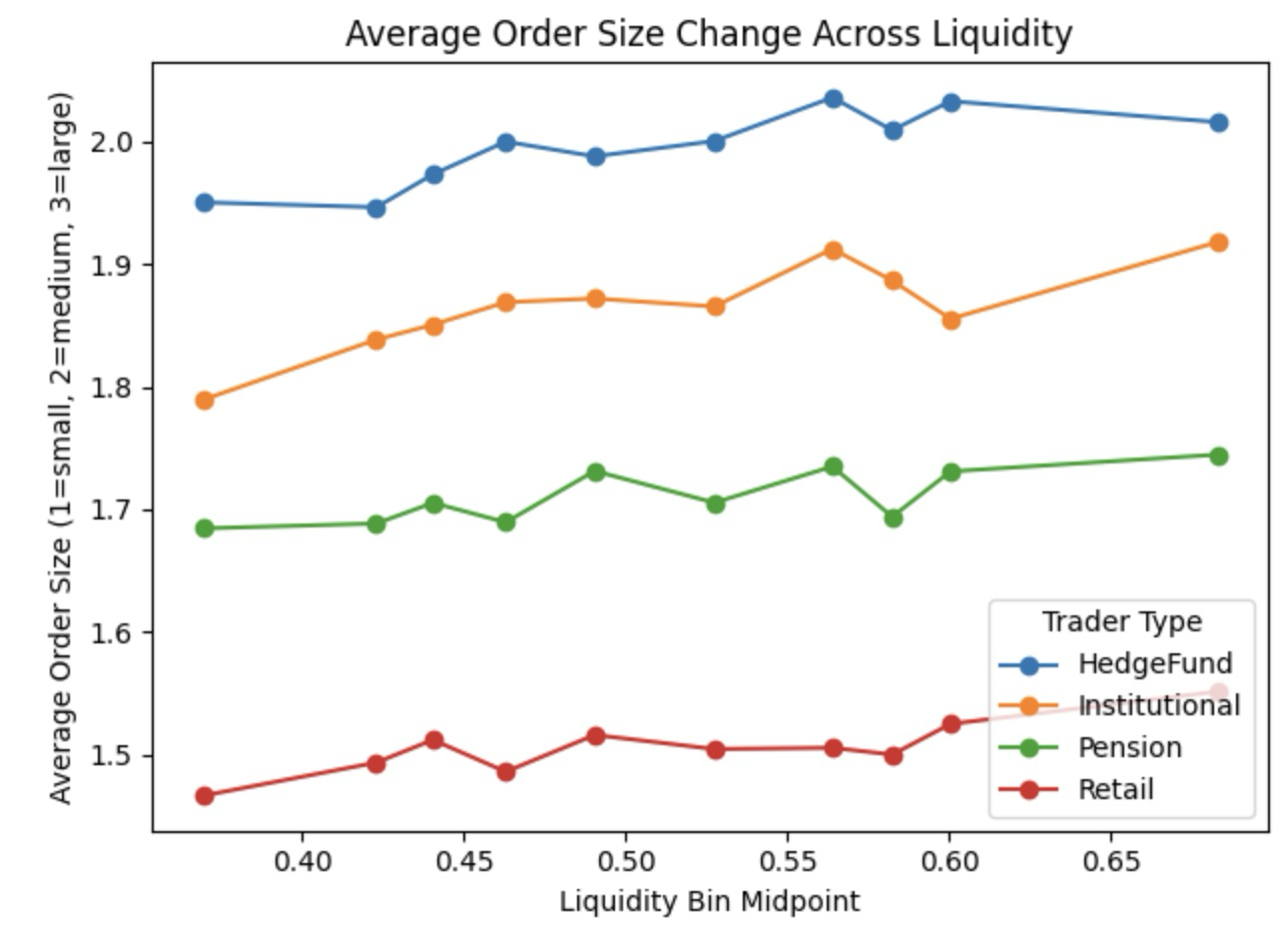}
    \caption{Order Size Under Different Liquidity}
    \label{fig:enter-label}
\end{figure}

\section{Discussion and Further Improvements}

The DGP is designed to model trader behavior under macroeconomic shocks, but it relies on simplifying assumptions that may not fully capture real-world complexities. Below, we outline several key limitations and potential refinements.

\textbf{Simplified Two Assets Structure}
The model includes only a risk-free asset and an inflation-exposed risky asset, assuming independence between them. Real markets feature diverse range of  assets with varying inflation sensitivities and complex covariance structures. Future work should incorporate a richer asset space with dynamic correlations and heterogeneous risk profiles.

\textbf{Static Risk Aversion}
We assume traders have fixed risk aversion levels, however, in reality, risk preferences is dynamic and change with market conditions. For instance, retail traders may become more risk-averse after getting large returns, while hedge funds may adjust risk-taking in response to what they perceive as structural market shifts. Future research should model dynamic risk aversion based on market feedback and trader adaptation instead of settle with constant risk aversion assumption.

\textbf{Return Distribution Assumptions}
We assume risk-free asset has a constant \(R_f = 4\%\) and that risky asset returns follow a normal distribution with a fixed mean 0.1\% and varying standard deviations based on trader information levels. (eg. Hedge Funds will be 0.25\% std, while other traders will have a higher std to adjust for their information level). However, in reality, many financial assets like QQQ (ETF for S\&P) has a return that exhibit skewness, fat tails, and volatility clustering. A more precise specification of asset types and return distributions is necessary to better capture market dynamics.

\textbf{Parameter Tuning}

Several parameter in our model, like the parameter within trader's utility function in deciding order size selection, are arbitrarily set with economic intuition behind aiming to approximate reality. 

For instance, to capture liquidity effects, we assume large orders yield significantly lower utility during low-liquidity periods. Liquidity is modeled as a normally distributed random variable:
\(\text{liquidity}_{\text{arr}} = 0.5 + 0.1 \cdot \mathcal{N}(0,1)\) to simulate period with different level of liquidity. And the surprise is also generated as a scaled version of a normally distributed random variable, which can be better modeled. Similarly after literature review, here is a parameter setup for trader heterogeneity to approximate the real market conditions.

\[
\begin{array}{|c|c|c|c|c|}
\hline
\text{Trader Type} & \text{Risk Aversion} & \text{Info Level} & \text{Max Risk} & \text{Base Transaction Cost} \\
\hline
\text{Retail} & 3.0 & 0 & 0.25 & 0.1 \\
\text{Pension} & 2.0 & 0.7 & 0.4 & 0.005 \\
\text{Institutional} & 1.5 & 0.8 & 0.8 & 0.005 \\
\text{HedgeFund} & 1.0 & 1.0 & 1.0 & 0.002 \\
\hline
\end{array}
\]

\textbf{Shock Model}

Currently, macroeconomic shocks are simulated as a combination of temporary and permanent components, where temporary shocks follow a mean-reverting normal process and permanent shocks follow a random walk. While this provides a useful decomposition, there are several areas for improvement:

\[
\begin{aligned}
    \text{temp\_shocks} &= 0.1 \cdot \mathcal{N}(0,1)_{T} \quad \text{(mean-reverting noise)} \\
    \text{perm\_shocks} &= \sum_{t=1}^{T} 0.05 \cdot \mathcal{N}(0,1)_t \quad \text{(random walk)} \\
    \text{cpi\_surprises} &= \text{temp\_shocks} + \text{perm\_shocks} 
\end{aligned}
\]

Currently, macroeconomic shocks are simulated as a combination of temporary and permanent components, where temporary shocks follow a mean-reverting normal process and permanent shocks follow a random walk. However, here we assume shocks are independent and identically distributed (iid) normal shocks. But in real life we should consider about relationship between shocks and use historical macroeconomic data to calibrate the persistence and distributional properties of shocks.\\

\end{document}